%% file: Hamil1_csqcd4.tex
\documentclass[12pt]{article}
\usepackage{epsfig}
\usepackage{graphicx}

\input econfmacros-csqcd4.tex
\def\Title#1{\begin{center} {\Large {\bf #1} } \end{center}}

\begin{document}

\Title{Braking index of isolated pulsars: open questions and ways forward}

\bigskip\bigskip


\begin{raggedright}

{\it 
Oliver Hamil$^{1}$\\
\bigskip
$^{1}$Department of Physics and Astronomy, University of Tennessee, Knoxville, Tennessee 37996 USA\\
}

\end{raggedright}

\section{Introduction}
Isolated pulsars are rotating neutron stars with accurately measured angular velocities $\Omega$, and their time derivatives which show unambiguously that the pulsars are slowing down. Although the exact mechanism of the spin-down is a question of debate in detail, the commonly accepted view is that it arises through emission of magnetic dipole radiation (MDR) from a rotating magnetized body. Other processes, including the emission of gravitational radiation, and of relativistic particles (pulsar wind), are also being considered. The calculated energy loss by a rotating pulsar with a constant moment of inertia is assumed proportional to a model dependent power of $\Omega$. This relation leads to the power law $\dot{\Omega}$ = -K $\Omega^{\rm n}$ where $n$ is called the braking index. The MDR model predicts $n$ exactly equal to 3. Selected observations of isolated pulsars provide rather precise values of $n$, individually accurate to a few percent or better, in the range 1$ <$ n $ < $ 2.8, which is consistently less than the predictions of the MDR model. In spite of an extensive investigation of various modifications of the MDR model, no satisfactory explanation of observation has been found as of yet.

\section{Braking Index}
The pulsar spin-down rate is an observation that has been recorded for the past few decades.  That pulsars can be seen and measured is evidence of the electromagnetic radiation emanating from what is assumed to be a strong dipole misaligned with the axis of rotation.  This misaligned dipole is generally considered to be the source of the pulsar spin-down as it accounts for rotational energy being carried away from the star \cite{pacini1967, pacini1968, gold1968, gold1969,goldwire1969}.  The observed loss of energy can be modeled by the simple power law, $\dot{\Omega}$ = -K $\Omega^{\rm n}$ where $n$ is the so-called the braking index.  The braking index itself is a purely observational parameter which is determined from pulsar timing observations along with their first and second derivatives \cite{lyne2015}
\begin{eqnarray}
n& = & {{{\Omega} \ddot{\Omega}} \over {{\dot{\Omega}}^2}}. 
\end{eqnarray}

The value for the braking index $n$ is theoretically determined by the torque mechanism working against the rotation of the star.  In the simple MDR model, the radiating dipole carries energy away from rotation thereby producing a torque which slows the rate of rotation of the pulsar.  This is the main assumption for pulsars because of the nature of the observed radiation, but there are at least two other acceptable possibilities for the torque mechanism.  The emission of charged particles, accelerated to relativistic velocities, forming a massive wind from the surface of the pulsar \cite{michel1969, harding1999} may play a role in spin-down.  Another consideration is higher order multipole electromagnetic radiation, or gravitational quadropole components to the radiated energy \cite{ostriker1969,alford2014}.  For each mechanism described above, the theoretical braking index is $n = 1,3,5$ for particle wind, MDR, and quadrupole radiation respectively.

\subsection{Problems}
Of the mechanisms described above, the most readily accepted is the MDR model.  This model predicts the radiated energy expected from a magnetized sphere rotating in vacuum (the accepted description of pulsars).  The braking index value for the MDR model is $n = 3$.  We see in Table~\ref{tab1} eight pulsars with accurately observed spin evolution.  The given values for the braking index are accurate to within a few percent or better.  It is clear that none of these observed stars has a braking index consistent with any of the values produced from theory.

\begin{table}[htb]
\centering{}%
\begin{tabular}{|c|c|c|c|}
\hline 
PSR                           &       Frequency          &  $n$                             & Ref.  \\
                                     &            (Hz)               &                                 &     \\ \hline
 B1509$-$58              &     6.633598804    &    2.839$\pm$0.001          &  \cite{livingstone2007}  \\
 J1119$-$6127          &     2.4512027814   &   2.684$\pm$0.002          & \cite{waltevrede2011}   \\
 J1846$-$0258          &     3.062118502     &   2.65$\pm$0.1                &\cite{livingstone2007}     \\
                                &                             &   2.16$\pm$0.13                             &\cite{livingstone2011} \\
 B0531+21 (Crab)      &    30.22543701      &   2.51$\pm$0.01               &  \cite{lyne1993} \\
 B0540$-$69             &    19.8344965        &   2.140$\pm$0.009            & \cite{livingstone2007,boyd1995}  \\
 J1833$-$1034         &    16.15935711      &   1.8569$\pm$0.001       & \cite{roy2012}   \\
 B0833$-$45 (Vela)    &    11.2                   &   1.4$\pm$0.2                   & \cite{lyne1996}   \\
 J1734$-$3333          &      0.855182765    &    0.9$\pm$0.2                &\cite{espinoza2011}   \\  \hline 
\end{tabular}\medskip{}
 \caption{Selected pulsars adopted from \cite{magalhaes2012,espinoza2011,lyne2015}. 
\label{tab1}}
\end{table}

There is a clear deviation of actual braking index values from the calculated ones.  The static value associated with each torque mechanism (MDR, wind, quadrupole) is directly due to the treatment of the differential equation governing braking index.  Each mechanism has the differential form,
\begin{eqnarray}
\dot{E} &=& -C{\Omega}^{n+1}, 
\end{eqnarray}
where C contains the physics of the associated mechanism, $\Omega$ is the rotational velocity, and $n$ is the braking index.  Likewise, the rotational energy of a rotating sphere is given by,
\begin{eqnarray}
\dot{E} &=& {d \over {dt}}({1 \over 2}I{\Omega}^2),
\end{eqnarray}
where $I$ is the moment of inertia.  Setting the above two equations equal to each other leads to two different treatments of the result.  Firstly, we can assume the moment of inertia is constant in time which leads to the known power law,
\begin{eqnarray}
\dot{\Omega} &=& -K{\Omega}^n,
\end{eqnarray}
where $K = C/I$.  This leads to the given values described earlier of $n = 1,3,5$.  A second approach is to consider that the moment of inertia changes in time which leads to braking index as a function of $\Omega$,
\begin{eqnarray}
n(\Omega) = n_0 - {{3\Omega I' + {{\Omega}^2}I''} \over {2I + \Omega I'}} + {{C'\Omega} \over C},
\label{eqn:freq}
\end{eqnarray}
where $n_0$ is the theoretical braking index associated with each torque mechanism, and the prime notation denotes derivatives with respect to angular frequency.

As we see from the above equations, the simple values for braking index come from assumptions that the moment of inertia of the star is constant (or that the star is static), and the parameters defining the physics of the braking torque are all constant.  This means that composition, magnetic field, rotational effects, etc. have no effect on braking.  The frequency dependent solution (Eq.~\ref{eqn:freq}) allows for the problem to be explored in much greater detail.  The microscopic proterties of the star must be considered in order to have a cohesive model which allows for changing values which may affect the braking index.

\section{Moving Forward}
\begin{figure}[htb]
\includegraphics[width=0.6\textwidth]{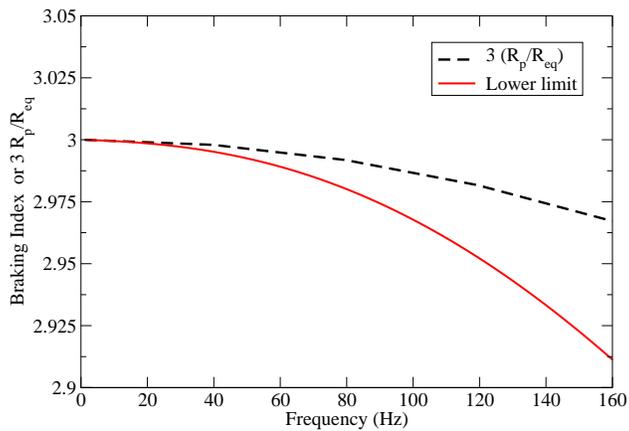}
\caption{Braking index as a function of frequency (solid line) compared with the ratio between polar (R$_{\rm p}$) and equatorial radii (R$_{\rm eq}$), normalized to three, which determines deformation of the star. The difference between the two lines represents a corelation between deviations of the braking index from  $n = 3$ and deformation for a 1.0 M$_{\odot}$ pulsar rotating at frequencies below 160 Hz (notice the expanded y-scale). It is seen that the shape deformation, even for this most deformable star (i.e. low mass, soft EoS), is small at these frequencies and quite unable to reproduce the observed range of braking indices.}
\label{fig10}
\end{figure}

The frequency dependent solution to the braking index allows for the problem to be approached from two extreme positions.  There can be a dependence of frequency where the moment of intertia changes at a rate which brings the braking index away from the canonical value, or in the limit of very slow rotation, the parameters governing the torque mechanism can be changing with frequency (or time).  Close examination of Eq.~\ref{eqn:freq} shows that changing moment of inertia can have a significant effect on the braking index at high frequencies, and in the limit that moment of inertia changes very slowly with frequency, there is still a term allowing for the physics in the torque mechanism ($C$) to change.  In the assumed MDR model, high frequency calculations show a reduction in braking index consistent with observation; however, the observed pulsars are all rotating at low frequency.  It is shown in Figure~\ref{fig10} the effect of changing moment of inertia over a range of frequency up to 160~Hz assuming constant $C$.  From the figure it is clear that below this frequency, the change in moment of inertia is negligible, and the subsequent deviation from $n = 3$ is also insignificant.  Noting that the data in Table~\ref{tab1} spans a range of frequencies between about 1 - 30~Hz, it is worthwhile to consider changes in the physical parameters of the torque mechanisms at low frequency.  

\subsection{Mechanism}
It is important to model pulsar braking indices for frequencies consistent with the accepted observations.  This can be approached in a few different ways.  The first thing to look at is the dependence on $C$ in Eq.~\ref{eqn:freq}.  For example, in the MDR model we have,
\begin{eqnarray}
C \propto {\mu}^2 \sin{\alpha}^2,
\end{eqnarray}
where $\mu$ is the magnetic moment and $\sin{\alpha}$ is the angle of inclination between the magnetic moment and the axis of rotation.  It is clear from Eq.~\ref{eqn:freq} that one or both of these values would have to increase as the pulsar spins down to move toward the observed braking index values that are less than $n=3$.  It has been suggested that there is some observatinal evidence that $\sin{\alpha}$ is increasing with time \cite{lyne2015}.

Effects on the magnetic field may be a factor for all three accepted torque mechanisms.  The magnetic field may increase, decrease, or change alignment \cite{contopoulos2006}.  The torque may vary with magnetic field in a way that is not a pure dipole.  Plasma outflow may cause currents in the magnetosphere.  Interaction with the magnetosphere in general may play in the torque mechanism \cite{livingstone2011}.  Rotationally driven effects such as a phase transition, density profile, and superfluidity may also affect the magnetic field.  The evolution of the magnetic field is most readily applied to the MDR model, but it is also important for the relativistic wind.

Particles near the surface of the star can be accelerated to relativistic energies, and thus carry away rotational energy from the pulsar.  The wind mechanism has a dependence on the magnetic field, and thus all of the considerations explained above can affect the braking index due to the wind.  The braking index value for the wind is $n=1$ which is near the lowest measured braking indices shown in Table~\ref{tab1}.  Since all values in Table~\ref{tab1} fall between $1$, and $3$, it is prudent to consider the wind as a possible torque mechanism affecting braking.

\subsection{Polynomial}
The considerations outlined above require knowledge of the origin and distribution of the magnetic field which is not readily available.  Also the nature of rotationally driven effects is not well known.  This poses a problem in that there is some speculation involved in trying to relate braking index values to changes in the magnetic field or other similar physics in the torque mecahnisms which leads to a phenomenology that is not necessarily physical.

Another approach to this problem considered by Alvarez et. al. \cite{alvarez2004} is to expand the braking law itself into a polynomial which consists of all assumed torque mechanisms.  In this way, we can assume there are no changes in the magnetic field, composition of the star, magnetosphere, etc., and that the pulsar is rotating at low frequency.  Given that the accurately measured braking indices range in value from about 1 - 2.8, it follows that the braking comes form a combination of torque mechanisms.  

As shown in \cite{alvarez2004} the braking law can be expanded as,
\begin{eqnarray}
\dot{\Omega} &=& -s(t)\Omega -r(t){\Omega}^3 -g(t){\Omega}^5,
\end{eqnarray}
where $s(t)$, $r(t)$, and $g(t)$ are functions representing the wind, MDR, and quadrupole torque mechanisms respectively, and $\Omega$ is the rotational frequency.  

The polynomial can be used to fit the known braking index values.  The braking index range of roughly $1-3$ indicates that the combination of wind ($n=1$) and MDR ($n=3$) should be important.  The quadrupole ($n=5$) may not play a role at low frequencies.  Furthermore, the combination of $s(t)$ and $r(t)$ may constrain the magnetic field.  The polynomial solution may also be expanded into high frequencies which may give insight into other mechanisms at work such as mass, composition, magnetic field, etc.

\section{Summary}
The problem of braking index can be approached from two ends.  We can consider very fast rotation where moment of inertia may dominate the spin evolution, or we can consider very slow rotation where the physics of the braking mechanism must be adjusted to fit the data.  The braking index is dominated by the torque mechanisms at frequencies below about $200Hz$.  Above this frequency, changes in moment of inertia due to deformation of the star become increasingly important.  Unknown composition and magnetic field dynamics play a crucial role in understanding the braking index, and are important to magnetic dipole radiation and relativistic wind at low frequency.  These effects are difficult to model as they are not yet well understood.

In order to model braking indices without knowledge of magnetic field evolution or rotationally driven changes in composition, we can construct a polynomial which will fit that data using functions of the known braking index mechanisms at low frequency.  This model can be extended to high frequencies, and to include considerations of the unknown values described above.  This can also be extended to include physical equations of state, and mass dependence.

A single, cohesive investigation of all the above possibilities is necessary to our understanding of pulsar spin evolution.  The resulting parameters may help to constrain the poorly known micro-physics at work in the cores of rotating neutron stars.



\subsection*{Acknowledgement}
We express our thanks to the organizers of the CSQCD IV conference for providing an 
excellent atmosphere which was the basis for inspiring discussions with all participants.

\end{document}

%% file: econfmacros-csqcd4.tex
\def\beq{\begin{equation}}
\def\eeq#1{\label{#1}\end{equation}}
\def\eeqn{\end{equation}}

\def\beqa{\begin{eqnarray}}
\def\eeqa#1{\label{#1}\end{eqnarray}}
\def\eeqan{\end{eqnarray}}

\let\bar=\overbar

\def\Dslash{\not{\hbox{\kern-4pt $D$}}}
\def\dslash{\not{\hbox{\kern-2pt $\del$}}}

\def\msb{{\bar{\ssstyle M \kern -1pt S}}}

\usepackage{fancyhdr,graphicx}